\documentstyle[seceq,epsf]{ptptex}
\def\tit#1#2#3#4#5{{#1} {\bf #2}, #3 (#4)}

\def\prl{Phys.\ Rev.\ Lett.\ }

\def\prb{Phys.\ Rev.\ B}

\title{%        %You can use \\ for explicit line-break
Resonating valence bond liquid physics on the triangular lattice
}
\author{%       %Use \sc for the family name
R. {\sc Moessner} and S. L. {\sc Sondhi}$^{*}$
}

\inst{%         %Affiliation, neglected when [addenda] or [errata]
Laboratoire de Physique Th\'eorique de l'Ecole Normale Sup\'erieure;\\
CNRS-UMR 8541;
24, rue Lhomond; 75231 Paris Cedex 05; FRANCE\\
$^*$ Department of Physics; Princeton University; Princeton, NJ 08544; USA
}

\recdate{%      %Editorial Office will fill in this.
}

\abst{We give an account of the short-range RVB liquid phase on the
triangular lattice, starting from an elementary introduction to
quantum dimer models including details
of the overlap expansion used to generate them. The fate of the topological 
degeneracy of the state under duality is discussed, as well as recent 
developments including its possible relevance for quantum computing.}

\begin{document}

\maketitle

\section{Short-range RVB physics --- some history}
The question posed by high-temperature superconductors is what happens
to an antiferromagnetic Mott insulator upon doping with mobile charge
carriers. The basic problem arises from the fact that a hole hopping
through an ordered N\'eel background is frustrated: in the absence of
spin flips, the hole leaves behind a 
trail of broken bonds; optimising
kinetic and exchange energy at the same time appears impossible.

Early on, Anderson\cite{pwa87} suggested that one way of resolving
this dilemma would be for the magnet to enter a state which has no
long-range order but is nonetheless energetically competitive. Such a
quantum state could be based on singlet bonds (valence bonds), as, for
a given pair of spins, these optimise the exchange energy. However,
this is again a frustrated problem as each spin can only form a
singlet bond with one of its neighbours. Hence, a product of singlet
wavefunctions has a hard time competing against the N\'eel state,
where each site gains energy from its four neighbours.

This energy can be retrieved, at least in part, by resonance processes
between different pairings of the spins into singlet bonds, as quantum
mechanically a spin can after all be in a superposition of singlet
bonds with different neighbours. One would then hope that such a sea
of singlet pairs would present less of an obstacle to hole motion
than the N\'eel state.

\begin{figure}
\epsfxsize=5in
\centerline{\epsffile{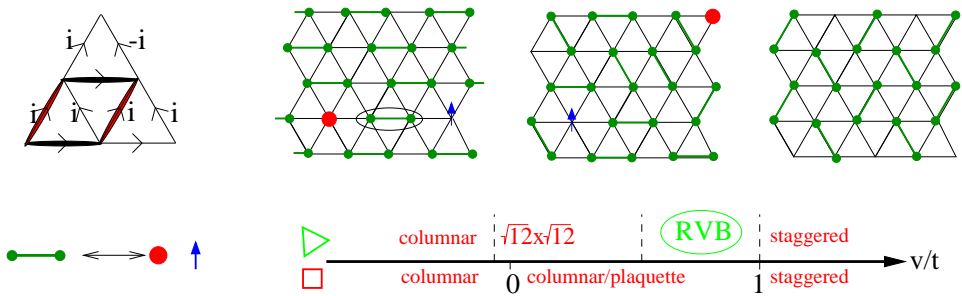}}
\caption{ Top left: The basic dimer resonances consist of replacing
the horizontal pair of dimers by the slanted pair, and the symmetry
equivalent moves. This process has kinetic energy $t$, and the
presence of such a flippable pair costs potential energy $v$.  The
Sutherland singlet sign convention for the triangular lattice is
indicated by arrows. Multiplying dimers on links by the phase factors
$\pm$i changes the sign of $t$. Bottom left: Knocking out an electron
leaves behind a hole and an upaired spin. Top right: Cartoons of
columnar, RVB and staggered phases. Whereas a hole hopping through the
columnar state generates a domain wall (encircled dimer), no such
impediment to hopping exists in the liquid phase. Bottom right: phase
diagram of square and triangular RK-QDM at zero temperature.}
\label{fig}
\end{figure}

Following this route, Kivelson, Rokhsar and Sethna developed what has
become known as short-range resonating valence bond (SR-RVB)
physics.\cite{kivrokset} In particular, the Rokhsar-Kivelson (RK)
quantum dimer model (QDM),\cite{Rokhsar88} which this account is in
large part devoted to, restricts the wavefunctions to contain products
of valence bonds between nearest neighbours only, as opposed to the
longer-range valence bonds included in more general
approaches. Somewhat disappointingly, it was found that such a model
would again lead to a ground state with broken translational
symmetry,\cite{Rokhsar88,subirfinitesize} except at an isolated
critical point, with order not in the spin-spin but rather in a
correlation function of singlets. The resulting crystalline order
again impedes hole motion in a manner completely analogous to the
situation in the N\'eel state (see Fig.~\ref{fig}).

The short-range RVB liquid was thus a phase in search of a
Hamiltonian.  In the following, we show that the RK quantum dimer
model on the {\em triangular} lattice, has an RVB liquid phase for a
finite region of parameter space, flanked by a number of crystalline
phases. Unlike those, the liquid is topologically ordered and exhibits
fractionalised deconfined excitations. We show how the topological
degeneracy gets lost under the duality transformation to an Ising
model, and close by mentioning recent developments. We have tried not
to duplicate any technical content, which can be found in
Refs.~\cite{mstrirvb} (RVB liquid), \cite{TFIM} (with Premi Chandra on
transverse field Ising models) and \cite{IGT} (with E. Fradkin on
gauge theories).

\section{The Rokhsar-Kivelson quantum dimer model}
In this section, we give a description of the RK-QDM on the triangular
lattice, following the original formulation for the square
lattice\cite{Rokhsar88,steveunpub}. The basic starting point is the
restriction of the Hilbert space of the spins with $S=1/2$\ residing
on the sites of the lattice to those states which can be written as a
product of singlet pairs between spins on neighbouring sites. Such
singlet pairs are conveniently denoted by dimers, and the requirement
that each spin be in a singlet pair with exactly one of its neighbours
implies that the restricted Hilbert space can be labelled by {\em
hardcore} dimer coverings of the lattice that the spins reside on.

The first issue to worry about is whether such states are linearly
independent. The answer to the first question on the square lattice is
yes.\cite{chayes} This appears to be the case as well for a
sufficiently large triangular lattice.\cite{mambriniprivate} For
increasingly highly-connected lattices, this will certainly cease to
be the case, as can easily be seen for a quadruplet of mutually
interconnected spins, which have two global singlet states but allow
three dimer coverings.

Quite generally, different dimer coverings are not orthogonal. In
fact, the overlap between two dimer coverings can be determined as
follows. One can form what is known as a transition graph by
superimposing two dimer configurations. Where the locations of a pair
of dimers coincides in the two coverings, one obtains a doubly
occupied bond. The other dimers give rise to closed loops, of varying
length $L_i$, in the transition graph. The overlap between the two
wavefunctions is the product over such loops of $2 x^L_i$, where
$x=1/\sqrt{2}$. 

\subsection{The overlap expansion}

Let us label the dimer coverings by $\left|i\right>$, so that their
overlap matrix (which is real and symmetric: $S=S^T$) is given by
$S_{ij}=\left<i\left|\right.j\right>$.  Let us now assume that
$S^{-\frac{1}{2}}$, the inverse square root of $S$,
exists.\cite{fn-sym} We define an orthonormal basis set
$\left\{\left|\alpha\right>\right\}$\ via
$\left|\alpha\right>=\sum_iS^{-\frac{1}{2}}_{\alpha,i}\left|i\right>$,
so that $ \left<\alpha\left|\right.\beta\right>=
S^{-\frac{1}{2}}SS^{{-\frac{1}{2}}^T}$, the unit matrix.  The matrix
elements of the antiferromagnetic nearest neighbour Heisenberg $S=1/2$
Hamiltonian $H=J\sum_{\left<ab\right>}\mathbf{S_a\cdot S_b}$ are hence
$\left<\alpha\left|H\right|\beta\right>=
\sum_{ij}S^{-\frac{1}{2}}_{i\alpha}\left<i\left|H\right|j\right>
S^{-\frac{1}{2}}_{j\beta}$. The trick devised by RK is to use the
parameter $x$\ defined above to obtain an expansion of this expression
which is then truncated.\cite{fn-largeN} One finds
$S^{-\frac{1}{2}}=1-x^4 \Box + O(x^6)$, where $\Box$\ equals one for
the matrix element of two dimer configurations differing only by a
single loop of length four in the transition graph, and vanishes
otherwise. Note that each orthogonal state can still be labelled by
the dimer covering, of which it contains an amplitude of $O(1)$.
Similarly, $\left<i\left|H\right|j\right>=
-\frac{3}{4}J\hat{n}_{d}-(\eta-\zeta-2)x^4\Box+O(x^6)$. Here, $\eta\
(\zeta)$\ are the number of bonds linking sites separated by an odd\
(even) number of steps along the loop in the transition graph.

To obtain the lowest order contributions to the QDM Hamiltonian,
$H_{RK}$, we first note that we are free to subtract a constant from
$H$\ as this only changes $H_{RK}$\ by the same constant. We thus
subtract the zeroth order term involving the number of dimers,
$-\frac{3}{4}J\hat{n}_{d}$, so that the effective Hamiltonian is given
by $H_{RK}=\left[1-x^4 \Box + O(x^6)\right]
\left[-(\eta-\zeta-2)x^4\Box+O(x^6)\right]\left[1-x^4 \Box +
O(x^6)\right]$, so that the lowest order term is a kinetic
(off-diagonal) one, namely $-(\eta-\zeta-2) x^4 \Box\equiv -t
\Box$. Its strength is thus given by $t=2 x^4$\ for the square
and $t=x^4$\ for the triangular lattice. 
%(In passing, we note the
%gratifying feature that at this order, the resonance term for spins on
%a tetrahedron, where $\eta=2\zeta=4$, vanishes.) 

We have so far ignored the issue of the sign of $t$. For the square
lattice, RK noted that the sign of $t$\ is a matter of convection, and
we can show that this remains the case for the triangular lattice. We
do this in two steps. Firstly, one can show by explicit computation
that the Sutherland phase convention for the square lattice can be
extended to the triangular lattice by choosing the sign of a singlet
bond involving sites $a,b$ such that
$(a,b)=[\uparrow_a\downarrow_b-\uparrow_b\downarrow_a]/\sqrt{2}$\ if
site $a$\ is below site $b$ or directly to its left. In this case, one
finds $t<0$.

To change the sign of $t$, one multiplies all basis states by a factor
of $i^{s_r + s_{l,g}-s_{l,u}}$. Here $s_l$\ counts the number of
dimers in that state on links pointing upwards and right, and
$s_{l,g}\ (s_{l,u})$\ count the number of dimers on links pointing
upwards and left from a site with an even (odd) vertical coordinate
(see Fig.~\ref{fig}). By this operation, every resonance move picks up
an overall minus sign, and hence $t\rightarrow -t$.

In the perturbation expansion, one generates a non-trivial diagonal
(potential) term at $O(x^8)$. This potential term (which is not
affected by the factors introduced above) counts the number of dimer
pairs which can participate in resonance moves, and has an amplitude
of $v=2(\eta-\zeta-2)x^8$. In the RK model, however, one treats
the ratio $v/t$\ as an adjustable parameter.

\section{The phase diagram --- the RVB liquid phase}
Two phases which occur as $v/t$\ is varied are easily identified. For
$v/t>1$, one finds that the ground state is one of twelve
symmetry-equivalent staggered states, which are annihilated by the
action of $H_{RK}$. For $v/t=-\infty$, there are $O(e^L)$\
degenerate states, but fluctuations appearing for finite $v/t$\ select
one (of six) columnar states at a high order in degenerate
perturbation theory. Whether any other crystalline phases, as
indicated in the phase diagram, are present
as $v/t$ is increased further is not settled.

The most interesting phase is a liquid anchored at the RK point
$v=t$,\cite{mstrirvb} where the diagonal correlations\cite{Rokhsar88}
of the QDM are identical to the correlations of the {\em classical}
dimer model. We have shown analytically that at the RK point, {\em
all} dimer correlations are short ranged. We have also argued that
this disordered point in fact terminates a gapped, liquidlike phase
which exists for a nonzero range of values $v/t<1$, and have provided
evidence from Monte Carlo simulations on large systems at low
temperature backing up this claim.\cite{mstrirvb}

This phase finally realises one important aspect of the short-range
RVB scenario: we have discovered a model, albeit an effective one,
that has an RVB liquid ground state. One can verify that this phase
bears all the important hallmarks, most crucially spin-charge
separation and holon deconfinement. If one knocks out one electron,
what is left behind is a hole on the site of the missing electron and
an unpaired spin on the site of its partner; in this model, both are
represented by monomers.  An easy way to see that these entities are
deconfined is provided by the following heuristic argument. Consider
the energy of two static monomers a distance $l$\ apart. If $l$\ is
much bigger than the correlation length, $\xi$, of the liquid, then
separating the monomers even further will not change this energy
significantly, as the effect of one monomer felt at the location of
the other is exponentially suppressed. The monomer potential thus
levels out and there is no confinement.

\section{Topological order and its fate under duality}

Although the liquid phase breaks no symmetries, there turn out to be
degenerate states of the QDM on a surface of nontrivial topology, for
example a torus. The origin of this degeneracy can be seen by forming
a transition graph of any state with some arbitrarily chosen reference
state. One then finds that the transition graphs can contain
noncontractible loops winding all the way around the torus; their
number is known as the winding number. Under any local dynamics, it
becomes impossible in the thermodynamic limit to have matrix elements
between dimer configurations with winding numbers differing by an odd
number. One thus obtains degenerate sectors, labelled `even' and
`odd'. This degeneracy in the absence of symmetry breaking is a
consequence of what Wen has termed `topological order',\cite{wenniu} a
concept we have critically discussed in this context in Ref.~\cite{IGT}.

{\noindent\bf Duality:} We have given a simple geometric demonstration
of a duality between generalised dimer models (formulated as an `odd'
Ising gauge theory,\cite{IGT} a topic of considerable current interest
following a recently proposed experiment by Senthil and Fisher for the
detection of fractionalisation\cite{SF} in the underdoped
cuprates\cite{bonn}) and fully frustrated Ising models in a transverse
field on the dual lattice. The basic idea is that each frustrated bond
in the fully frustrated Ising model (on the dual hexagonal lattice) is
denoted by a dimer; the frustration of the Ising model implies that
the number of dimers per site is odd. The condition of only one dimer
per site translates into considering only ground states of the Ising
model in zero field.  The RVB phase is dual to a $T=0$ paramagnetic
phase of the Ising model. As the paramagnetic phase is known to lack
any topological order while the RVB phase manifestly possesses such
order, it is of interest to understand how duality squares this particular
circle.

First, note that the mapping of spins onto dimers loses the Ising
symmetry: a globally Ising reversed state produces the same dimer
configuration. The Ising states can be paired into symmetric and
antisymmetric combinations of Ising reversed states, which form
disconnected sectors. The Hamiltonian in the former sector, which
contains the system's ground state, is identical to that of the dimer
model, whereas in the latter sector, where the lowest excited state
may reside, some matrix elements have the `wrong' sign.\cite{TFIM}
However, a gap in the symmetric sector (dimer model) will generally
imply a gap for the Ising system.

In the reverse direction, note that the the number of dimers crossed
by a noncontractible loop on the torus encodes the number of spin
flips encountered. For one (the other) sector, this number will be
even (odd). For a consistent assignment of Ising spins,
(anti-)periodic boundary conditions are thus required. Hence, the
topologically degenerate sectors of the QDM map onto Ising models with
different boundary conditions, none of which possess topological order
{\it on their own}. This particular situation is special to $2+1$
dimensions, where gauge theories are dual to spin systems, and should
not be taken to indicate anything other than a great convenience in
studying dimer models; e.g. in $3+1$ gauge theories are dual to gauge 
theories and both sets of variables would immediately suggest the 
possibility of topological order while leaving the difficulty of 
studying the system unaltered!

\section{Other realizations - quantum computing}

The mathematics of the dimer model described above has been found
useful in other, closely related problems in subsequent work.

Nayak and Shtengel\cite{naysht} have written down Ising models in
which the up/down spins live on links and correspond precisely to 
the presence/absence of dimers and the Hamiltonian is the transcription
of the QDM to the spin language. The excitations are somewhat different
though: their spinon is our holon and their holon corresponds in
the QDM to removing an empty bond surrounding a holon altogether---which
is not natural in the latter but makes perfect sense in the spin
interpretation. 

Along this theme, Balents, Fisher and Girvin \cite{bfg} have analysed
an Ising model on the kagome 
lattice with ring-exchange dynamics where the spins can again be
thought of as the presence or absence of dimers on the triangular
lattice centered on the hexagons of the kagome. For the Hamiltonian
in Ref~\cite{bfg}, the new feature is that there are now exactly
{\it three} nonoverlapping dimers per site, rather than just a single 
one. With the same Hamiltonian as considered above, one gets again an
RK point which has been shown numerically to be liquid. Translated
back into spin language this yields deconfined $S_z = 1/2$
excitations. 

{\bf \noindent Quantum computing:} Following the seminal suggestion of
Kitaev,\cite{kitaev} Nayak and Shtengel\cite{naysht} noted that
the degenerate liquid states of the kind we find in the triangular RVB
liquid could, in principle, be used to construct q-bits. These are 
hoped to be especially stable as
tunneling events are suppressed because the necessary matrix
elements are absent for topological reasons.

A concrete realisation of such a q-bit, using the triangular RVB
liquid, was very recently proposed via a realisation of the triangular
QDM in a Josephson junction array.\cite{joarray} This proposal
includes an idea of how to tune the tunneling rate between the
topologically degenerate states, essentially by considering a system
on an annulus near the inner, small circumference of which links can
be weakened so as to lower the energy cost for the formation of a
defect. On moving around the inner circumference, this defect induces
tunnelling, the rate being exponentially sensitive to the energy cost
of the defect. While the task of fabricating such a device is by
no means straightforward, it is encouraging that it looks almost
possible!

\section{Conclusion}
The QDM on the triangular lattice realises a bona fide RVB liquid
phase. This demonstrates, as a point of principle, the possibility of
constructing fractionalised phases via the SR-RVB scenario, as well as
providing a route between a microscopic model and an effective Ising
gauge theory. Its precise relation to other proposed fractionalised 
triangular phases \cite{sachdev3,misguich} is at this point unclear 
although their commonality in requiring the triangular lattice
indicates a close connection. Finally, there is the interesting
possibility that the liquid phase discovered by us may prove to be of 
use in the context of quantum computing.

%\section{References}

\section*{Acknowledgements}
Parts of the work reviewed in this talk were undertaken in
collaboration with Premi Chandra and Eduardo Fradkin. We would like to
thank Steve Kivelson for access to his unpublished notes. This work
was supported in part by grants from the Deutsche
Forschungsgemeinschaft, the NSF (grant No. DMR-9978074), the
A. P. Sloan Foundation and the David and Lucille Packard Foundation.

%\appendix
%\section{First Appendix} %Empty argument \section{} yields `Appendix'. 

%\section{Second Appendix}

\end{document}